\documentclass[runningheads]{llncs}
\usepackage{amsmath,amssymb,graphicx}
\usepackage{booktabs,array}
\usepackage[T1]{fontenc}
\usepackage[expansion=false,protrusion=false]{microtype}
\usepackage[hidelinks]{hyperref}
\hypersetup{
  pdftitle={Anchored Fusion for Automated Airway Segmentation in Pediatric and Adult CT},
  pdfauthor={Huseyin Umut Isik, Ege Mert Balcik, Duygu Atasoy, Asutosh Sahu, \c{S}eyda Ertekin, Sila Kurugol}
}
\newcommand{\amr}{AirMorph-R}
\newcommand{\afuse}{A-Fuse}
\begin{document}

\title{Anchored Fusion for\\Automated Airway Segmentation\\in Pediatric and Adult CT}
\titlerunning{Anchored Fusion for Airway Segmentation}

\author{Huseyin Umut Isik\inst{1,3,\star} \and
Ege Mert Balcik\inst{1,3,\star} \and
Duygu Atasoy\inst{3} \and
Asutosh Sahu\inst{3} \and
\c{S}eyda Ertekin\inst{1,2,\star\star} \and
Sila Kurugol\inst{3,\star\star}}
\authorrunning{H. U. Isik et al.}

\institute{Department of Computer Engineering, METU, Ankara, Turkey \and
METU-DTX Digital Transformation and Innovation Center, Ankara, Turkey \and
Department of Radiology, Boston Children's Hospital and Harvard Medical School, Boston, MA, USA}

\maketitle
{\renewcommand{\thefootnote}{}\footnotetext{$^{\star}$Equal contribution. $^{\star\star}$Equal contribution as senior authors.\\
This work has been submitted to the IEEE for possible publication. Copyright may be transferred without notice, after which this version may no longer be accessible.}}

\begin{abstract}
Airway segmentation from chest CT enables branch-level measurement of airway size and growth, but peripheral bronchi remain hard to segment, particularly in children, whose airways are smaller and for whom few annotated scans exist. Different models recover complementary branches of the airway tree. Their union improves overall tree coverage but also introduces disconnected non-airway objects. We propose anchored fusion (\afuse{}), which takes the voxel-wise union of TfeNet, a retrained MEDPSeg and a reimplemented AirMorph and retains the connected component with greatest overlap with the AirMorph mask. On 30 held-out adult CTs from the ATM\textquotesingle26 training set, \afuse{} reaches a Dice score of 93.24\% and the highest branch detection rate among compared methods (92.05\%). On nine pediatric scans, it reaches 94.15\% Dice and 68.79\% branch detection. \afuse{} retains the airway tree in all 39 test scans, whereas keeping the largest component selects a non-airway object in four. \afuse{} is deployed as an automated web application and we release all code, trained weights and an open-source ITK-SNAP extension for 3D tubular segmentation editing.

\end{abstract}

\section{Introduction}
Airway segmentation supports branch-specific measurements of airway structure, including changes associated with disease and growth~\cite{lv2025normative}. Reliable tree recovery is particularly important for peripheral branches: a missed branch is lost for further branch-level analysis, such as assessing airway growth and disease in children, even when whole-volume overlap remains high. Airway benchmarks therefore report branch detection and tree-length metrics alongside the voxel Dice score~\cite{atm22}.

Deep networks have improved airway extraction across varied populations and imaging conditions~\cite{garcia2021}, but most training data come from adults. Children have smaller airways and more variable inspiration levels, so adult-trained models miss more peripheral branches in their scans. State-of-the-art methods such as AirMorph, MEDPSeg and TfeNet, all trained on adult CT, differ in architecture and training~\cite{zhang2025airmorph,carmo2024medpseg,wu2025tfenet}. Their predictions are often complementary, which matters most in children, where each model misses different branches. Union fusion preserves a branch detected by only one model, whereas majority voting and label-fusion methods such as STAPLE~\cite{warfield2004staple} weight agreement between models and may discard it. However, union also keeps model-specific false positives, many of them disconnected from the airway.

Disconnected false positives are commonly removed by keeping the largest connected component (LCC)~\cite{atm22}, as in the MEDPSeg and TfeNet pipelines. This assumes that the airway is the largest object in the mask, which fails when a non-airway object is larger, as in the infant scan in Fig.~\ref{fig:components}. We therefore select the airway component by its overlap with the   AirMorph mask rather than by its size. This keeps the fusion robust on pediatric CT without requiring pediatric training.

\begin{figure}[!h]
\centering
\includegraphics[width=1.0\textwidth]{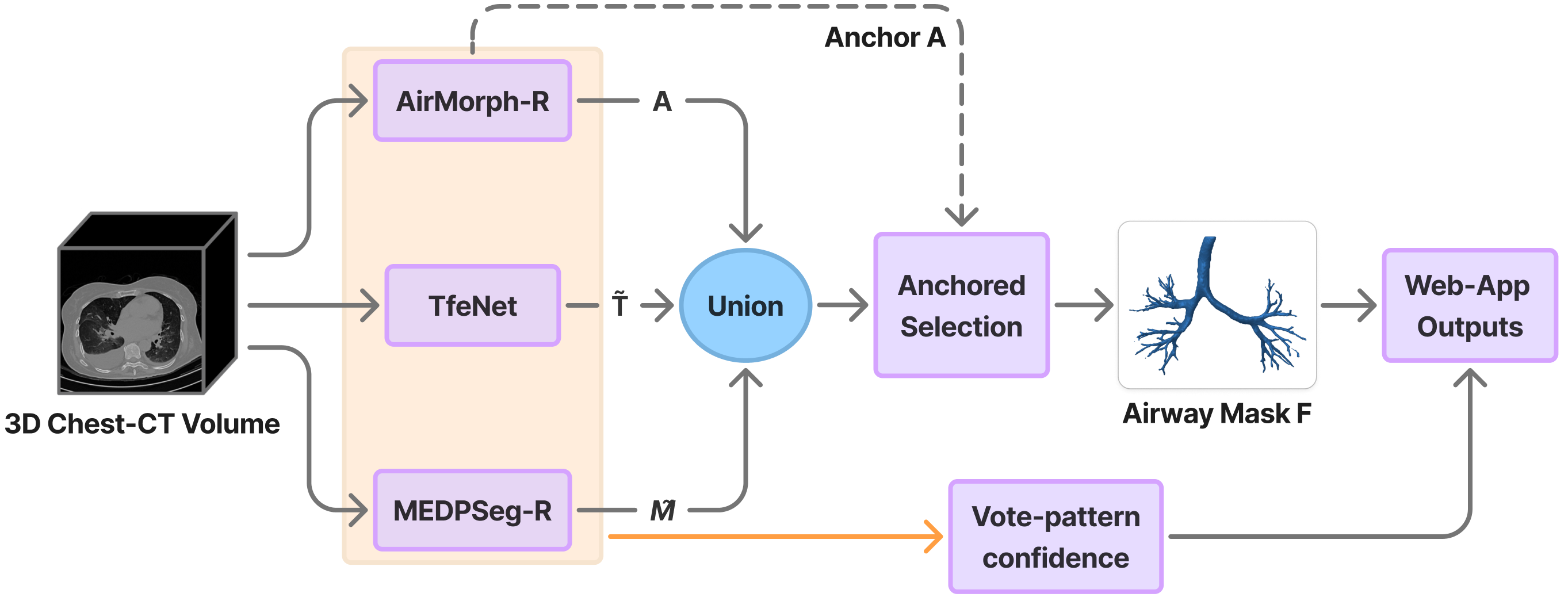}
\caption{Overview of our method. Segmentation (top to bottom): \amr{} produces $A$; MEDPSeg-R an    d TfeNet contribute their raw outputs $\tilde M$ and $\tilde T$. Their union $U=A\cup\tilde M\cup\tilde T$ is reduced to the connected component with greatest overlap with $A$, yielding the airway mask $F$ (Eq.~\ref{eq:anchor}). Confidence (right): each voxel's prediction pattern across \amr{}, MEDPSeg-R, and TfeNet is mapped to its precision on the adult validation set, providing voxel-level and branch-level confidence without modifying $F$.
}
\label{fig:architecture}
\end{figure}

\noindent\textbf{Our contributions are:}

\noindent\textbf{Three-model fusion with anchored airway selection.}
The voxel-wise union of TfeNet, MEDPSeg and AirMorph is reduced to the connected component that overlaps the AirMorph mask most, avoiding the failures of largest-component selection.

\noindent\textbf{Adaptation to pediatric CT.}
Existing adult-trained models may miss peripheral branches in children. \afuse{} outperforms these models in Dice and branch detection without pediatric training and retains the airway tree where largest-component selection fails.

\noindent\textbf{An open-source ITK-SNAP extension.}
It is developed for simple 3D editing of segmentation of tubular structures such as airways and was used to generate
pediatric ground truth segmentations.

\noindent\textbf{A fully automated web application.}
It returns the airway mask, centerline and branch-level measurements of an uploaded CT scan.
\section{Methods}
\subsection{Segmentation members and component selection}
\afuse{} combines three segmentation models (Fig.~\ref{fig:architecture}). \amr{} is our reimplementation of the AirMorph ensemble~\cite{zhang2025airmorph}. Because its training code and checkpoints were unavailable, 
we reproduced the method described in~\cite{zhang2025airmorph} in PyTorch. We trained five networks, $A_1,\ldots,A_5$ (three 3D U-Nets and two WingsNets) using the topology-aware losses described in~\cite{zhang2025airmorph}. Training used 239 ATM'26 training volumes, excluding the 30 volumes reserved for testing and 30 for validation. Code and checkpoints are publicly available\footnote{\label{fn:repo}\url{https://github.com/quin-med-harvard-edu/airway-seg}}. The five networks are combined by a three-of-five voting followed by largest-connected-component extraction, yielding the \amr{} mask $A$. Voting suppresses most false positives, and the largest remaining component is the airway in all test scans. Because the released MEDPSeg weights were trained on ATM data that may overlap our test set, we retrained  MEDPSeg on the same 239-volume training partition (MEDPSeg-R). TfeNet uses its released checkpoint trained on the BAS dataset. We denote the post-processed MEDPSeg-R and TfeNet outputs by $M$ and $T$, and their raw outputs by $\tilde M$ and $\tilde T$.  Fusion requires only binary masks and does not require comparable probabilities. 

For the union $U=A\cup\tilde M\cup\tilde T$, let $\mathcal{C}_6(U)$ be its 6-connected components. The final segmentation is given by
\begin{equation}
F=\operatorname*{arg\,max}_{C\in\mathcal{C}_6(U)}|C\cap A|,
\label{eq:anchor}
\end{equation}
i.e. the component $C$ that overlaps the \amr{} mask most. We call this method \afuse{}. Selecting after the union lets one member reconnect a fragment recovered by another. The anchor ranks components and does not intersect the output with $A$, so connected extensions supplied by MEDPSeg-R or TfeNet remain in $F$. 

\subsection{Deployed web application}
\afuse{} is deployed as a web application that runs the evaluated fusion pipeline without user input. Its link is provided in our code repository\textsuperscript{\ref{fn:repo}}. From an uploaded de-identified CT volume, it returns the airway mask $F$, its centerline and per-branch measurements (length, diameter, branching angle and generation), displayed in an interactive 3D viewer and downloadable in the orientation of the input volume. Each voxel of $F$ is also colored by its vote-pattern confidence. Processing a $512\times512\times800$ volume takes about ten minutes on one serverless NVIDIA A10G GPU.
\begin{figure}[tb]
\centering
\includegraphics[width=\textwidth]{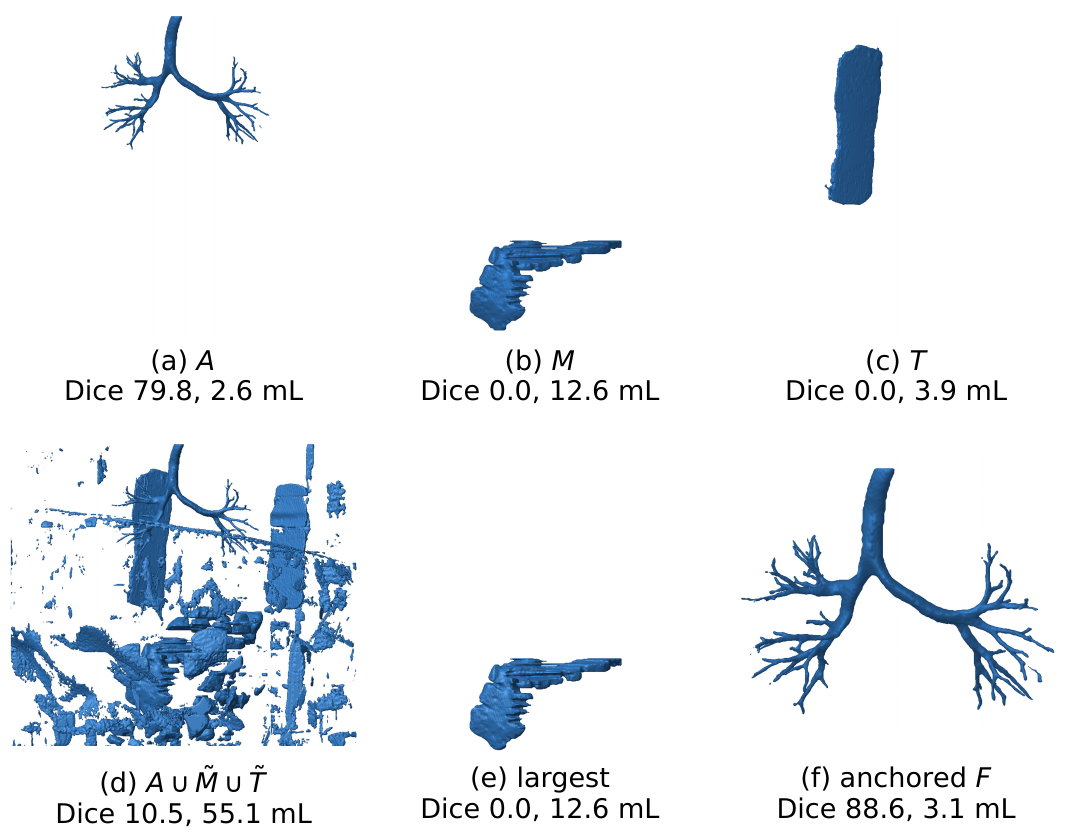}
\caption{Component selection in the 0.9-year-old infant. Top:  model outputs after their standard post-processing, (a) \amr{} ($A$), (b) MEDPSeg-R ($M$) and (c) TfeNet ($T$); (b) and (c) select non-airway components. Bottom: (d) union of $A$ with the raw outputs $\tilde M$ and $\tilde T$, (e) its largest component and (f) anchored selection $F$ (enlarged). Panels (a)--(c) and (e) share one scale. Dice against the ground truth and mask volume are given below each panel.}
\label{fig:components}
\end{figure}

\subsection{ITK-SNAP 3D editing extension}
We release a modified version of ITK-SNAP~\cite{itksnap2006}\footnote{\url{https://github.com/quin-med-harvard-edu/itksnap-3d-edit}} with added 3D segmentation editing tools, including painting and erasing in 3D, connected component deletion, tubular bridging of gaps between branch ends and local morphological closing. It runs locally and independently of the web application. It was used to create pediatric ground-truth segmentations (Section~\ref{sec:data}).

\section{Experimental Design}
\subsection{Cohorts and data separation}\label{sec:data}
Adult data comprise the 299 annotated training CT volumes of Track~1 of the ATM\textquotesingle26 challenge\footnote{\url{https://atm26.grand-challenge.org/}}, the successor of ATM\textquotesingle22~\cite{atm22}. We split them into 239 training, 30 validation and 30 test volumes. These are local partitions, not the official challenge test set. The five \amr{} networks and MEDPSeg-R were trained on the 239-volume partition, with the 30 test volumes reserved for evaluation.

Pediatric evaluation uses nine non-contrast chest CT examinations of eight patients aged 0.9 to 16.0 years from Boston Children's Hospital. The children were clinically free of respiratory disease and their CT examinations had previously been interpreted as normal. Slice interval was 0.3\,mm and in-plane resolution was 0.42--0.73\,mm. Board-certified pediatric radiologists created the ground-truth segmentations using our modified ITK-SNAP.

\begin{table*}[tb]
\centering
\caption{Segmentation results (mean; \% except HD95 in mm). $A$, $M$, $T$: outputs after each model's own post-processing; $\tilde M$, $\tilde T$: raw outputs. Anchored: component overlapping $A$ most; Ours: anchored $A\cup\tilde M\cup\tilde T$. Bold/underline: best/second-best per cohort.}
\label{tab:segmentation}
\fontsize{9}{9}\selectfont
\setlength{\tabcolsep}{2.5pt}

\resizebox{\textwidth}{!}{%
\begin{tabular}{lrrrrrrrrrrrr@{}}
\toprule
 & \multicolumn{6}{c}{Adult ATM\textquotesingle26 ($n=30$)} & \multicolumn{6}{c}{Pediatric CT ($n=9$)}\\
\cmidrule(lr){2-7}\cmidrule(lr){8-13}
Strategy & DSC & clDice & TLD & BND & Sens. & HD95 & DSC & clDice & TLD & BND & Sens. & HD95 \\
\midrule
AirMorph-R ($A$) & 92.06 & 90.29 & 94.30 & 90.81 & 94.92 & 2.32 & 87.70 & 87.85 & 76.10 & 59.81 & 78.27 & 2.80 \\
MEDPSeg-R ($M$) & \underline{93.84} & 87.59 & 79.29 & 68.38 & 93.16 & 2.11 & 78.94 & 59.39 & 41.63 & 29.51 & 71.14 & 21.27 \\
TfeNet ($T$) & 84.44 & 80.86 & 74.94 & 64.96 & 81.64 & 19.06 & 70.06 & 66.45 & 55.90 & 41.87 & 63.80 & 26.54 \\
\midrule
$A\cup M$ & 93.17 & 90.11 & 94.55 & 91.27 & \underline{97.84} & 1.80 & 86.64 & 86.66 & 77.63 & 61.01 & 86.53 & 12.37 \\
$A\cup T$ & 89.83 & 87.37 & \underline{95.03} & \underline{91.86} & 97.43 & 15.46 & 84.32 & 85.19 & 80.37 & 65.97 & 83.62 & 19.12 \\
$M\cup T$ & 90.59 & 86.21 & 86.71 & 78.08 & 95.30 & 15.36 & 79.23 & 72.73 & 62.49 & 46.97 & 76.92 & 24.26 \\
Majority Vote & \textbf{94.42} & 90.36 & 85.86 & 76.74 & 94.24 & \textbf{1.55} & 79.55 & 72.89 & 58.66 & 41.03 & 72.09 & 3.47 \\
\midrule
$A\cup\tilde M$, anchored & 93.21 & \underline{90.57} & 94.27 & 90.99 & 97.80 & \underline{1.72} & \underline{93.47} & 89.18 & 78.34 & 61.39 & \underline{87.97} & \underline{0.90} \\
$A\cup\tilde T$, anchored & 93.29 & \textbf{90.83} & 94.80 & 91.58 & 97.43 & 1.75 & 92.21 & \underline{90.79} & \underline{81.80} & \underline{68.15} & 85.72 & 1.30 \\
$A\cup\tilde M\cup\tilde T$, largest & 84.73 & 82.84 & 85.32 & 82.23 & 88.36 & 19.00 & 84.31 & 80.64 & 72.77 & 60.99 & 80.30 & 13.99 \\
\midrule
\textbf{A-Fuse (Ours)} & 93.24 & 90.48 & \textbf{95.09} & \textbf{92.05} & \textbf{98.21} & 1.77 & \textbf{94.15} & \textbf{91.11} & \textbf{82.52} & \textbf{68.79} & \textbf{89.22} & \textbf{0.70} \\
\bottomrule
\end{tabular}}
\end{table*}

\subsection{Comparisons}
We compare each model, all pairwise unions and majority voting of $A$, $M$ and $T$, without additional component selection. We also evaluate anchored fusion of  $A\cup\tilde M$ and $A\cup\tilde T$, and largest-component selection from $A\cup\tilde M\cup\tilde T$, to isolate the effects of fusion members and component selection. All variants use the same inputs.
\subsection{Metrics}
Segmentation metrics are Dice similarity coefficient (DSC), sensitivity, clDice~\cite{cldice}, tree length detected (TLD), branch detection (BND) and the 95th percentile Hausdorff distance (HD95, mm). DSC, sensitivity, TLD and BND are computed using the ATM'22 evaluation code~\cite{atm22}. TLD is the fraction of ground-truth centerline voxels contained in the segmentation. BND is the fraction of ground-truth branches with at least 80\% of their centerline contained in the segmentation. Metrics are computed per scan and averaged.

\section{Results}\label{sec:results}
\subsection{Segmentation and component selection}
In adults, \afuse{} reaches DSC 93.24\%, clDice 90.48\%, TLD 95.09\% and BND 92.05\% (Table~\ref{tab:segmentation}). Majority voting has higher DSC (94.42\%) and lower HD95 (1.55\,mm) but detects only 76.74\% of branches. The best pair for branch detection, $A\cup T$, reaches 91.86\% BND but retains disconnected non-airway objects, reducing DSC to 89.83\% and increasing HD95 to 15.46\,mm versus 1.77\,mm for \afuse{}. In nine pediatric examinations, \afuse{} performs best on all six metrics with DSC $94.15\pm2.2\%$ and BND $68.79\pm11.0\%$, 2.82 percentage points above the best plain pairwise union. In contrast, the members' own post-processing fails completely in some children (DSC SD 29.7 for MEDPSeg-R and 39.8 for TfeNet).  Using the same anchoring rule, $A\cup\tilde T$ and $A\cup\tilde M$ reach BNDs of 91.58\% and 90.99\% in adults and 68.15\% and 61.39\% in children. Adding the third model increases BND by 0.47 points in adults (95\% CI 0.31--0.64; $p<0.001$, paired Wilcoxon) and by 0.64 in children (6 of 9 better, none worse).

Anchored and largest-component selection differ in four cases: three of the 30 adults and the infant (Fig.~\ref{fig:components}). In all four, the largest component contains no airway. Across adults, DSC is $93.24\pm4.44\%$ with anchored selection versus $84.73\pm28.79\%$ with largest-component selection. The same failure occurs with the members' own largest-component post-processing: across all 39 scans, a non-airway component is selected in five cases for TfeNet, one for MEDPSeg-R (the infant, Fig.~\ref{fig:components}b) and none for \amr{}.   

\begin{table*}[t]
\centering
\caption{Adult results per generation (mean, \%). Generation 1 is the trachea; generations 8--10 are present in 29 of the 30 volumes. WT: whole tree. Members use their own post-processing, as in Table~\ref{tab:segmentation}. Bold indicates the best values for each metric and generation.}
\label{tab:generation}
\fontsize{9}{9}\selectfont
\setlength{\tabcolsep}{2.1pt}
\resizebox{\textwidth}{!}{%
\begin{tabular}{llrrrrrrrrrrr@{}}
\toprule
Metric & Strategy & 1 & 2 & 3 & 4 & 5 & 6 & 7 & 8 & 9 & 10 & WT\\
\midrule
DSC & AirMorph-R & 95.15 & \textbf{98.29} & \textbf{97.32} & 95.06 & 90.86 & 86.00 & \textbf{81.06} & \textbf{81.29} & \textbf{80.23} & \textbf{75.74} & 92.06 \\
 & MEDPSeg-R & 98.29 & 98.26 & 97.10 & \textbf{95.12} & \textbf{91.74} & \textbf{87.33} & 80.63 & 79.93 & 77.97 & 73.99 & \textbf{93.84} \\
 & TfeNet & 87.57 & 87.89 & 87.06 & 85.58 & 83.44 & 80.37 & 75.48 & 77.52 & 76.00 & 70.92 & 84.44 \\
 & \textbf{A-Fuse} & \textbf{98.44} & 98.13 & 96.98 & 94.72 & 90.46 & 85.62 & 80.50 & 80.88 & 79.76 & 75.27 & 93.24 \\
\midrule
clDice & AirMorph-R & 95.68 & \textbf{100.00} & \textbf{99.88} & \textbf{99.57} & 95.73 & \textbf{92.14} & 87.50 & 88.13 & 87.28 & 83.66 & 90.29 \\
 & MEDPSeg-R & 99.86 & 99.96 & 99.61 & 99.20 & 95.23 & 90.80 & 83.82 & 83.51 & 82.71 & 79.51 & 87.59 \\
 & TfeNet & 89.59 & 90.00 & 89.99 & 89.83 & 87.59 & 84.31 & 78.45 & 81.56 & 80.12 & 75.07 & 80.86 \\
 & \textbf{A-Fuse} & \textbf{99.97} & 99.96 & 99.57 & 99.48 & \textbf{95.77} & 91.91 & \textbf{87.82} & \textbf{88.28} & \textbf{87.92} & \textbf{84.05} & \textbf{90.48} \\
\midrule
TLD & AirMorph-R & \textbf{100.00} & \textbf{100.00} & \textbf{100.00} & 99.71 & 98.63 & 95.81 & 93.35 & 92.81 & 93.01 & 89.27 & 94.30 \\
 & MEDPSeg-R & \textbf{100.00} & \textbf{100.00} & \textbf{100.00} & 99.06 & 93.64 & 85.93 & 75.25 & 71.96 & 71.74 & 70.52 & 79.29 \\
 & TfeNet & 90.00 & 90.00 & 90.00 & 89.33 & 85.80 & 81.85 & 72.92 & 74.04 & 72.51 & 69.17 & 74.94 \\
 & \textbf{A-Fuse} & \textbf{100.00} & \textbf{100.00} & \textbf{100.00} & \textbf{99.79} & \textbf{98.92} & \textbf{96.60} & \textbf{94.27} & \textbf{93.60} & \textbf{93.96} & \textbf{90.26} & \textbf{95.09} \\
\midrule
BND & AirMorph-R & \textbf{100.00} & \textbf{100.00} & \textbf{100.00} & 99.03 & 98.47 & 96.00 & 89.87 & 87.93 & 87.68 & 82.42 & 90.81 \\
 & MEDPSeg-R & \textbf{100.00} & \textbf{100.00} & \textbf{100.00} & 98.61 & 93.76 & 83.67 & 66.18 & 57.47 & 55.18 & 53.81 & 68.38 \\
 & TfeNet & 90.00 & 90.00 & 90.00 & 88.47 & 86.18 & 79.53 & 63.29 & 62.27 & 58.66 & 53.18 & 64.96 \\
 & \textbf{A-Fuse} & \textbf{100.00} & \textbf{100.00} & \textbf{100.00} & \textbf{99.31} & \textbf{99.02} & \textbf{96.80} & \textbf{90.95} & \textbf{89.95} & \textbf{90.84} & \textbf{84.35} & \textbf{92.05} \\
\midrule
Sens. & AirMorph-R & 92.99 & 99.29 & 98.76 & 98.33 & 97.82 & 95.77 & 93.45 & 93.11 & 93.28 & 88.66 & 94.92 \\
 & MEDPSeg-R & 98.85 & 99.06 & 98.22 & 96.51 & 92.94 & 86.46 & 77.69 & 74.92 & 73.23 & 70.70 & 93.16 \\
 & TfeNet & 85.59 & 86.20 & 84.84 & 82.64 & 80.38 & 77.87 & 72.91 & 74.47 & 72.55 & 69.07 & 81.64 \\
 & \textbf{A-Fuse} & \textbf{99.52} & \textbf{99.46} & \textbf{98.96} & \textbf{98.71} & \textbf{98.20} & \textbf{96.46} & \textbf{94.15} & \textbf{94.45} & \textbf{94.26} & \textbf{89.66} & \textbf{98.21} \\
\bottomrule
\end{tabular}}
\end{table*}

\subsection{Generation-wise airway recovery}
\begin{figure}[!h]
\centering
\includegraphics[width=0.9\textwidth]{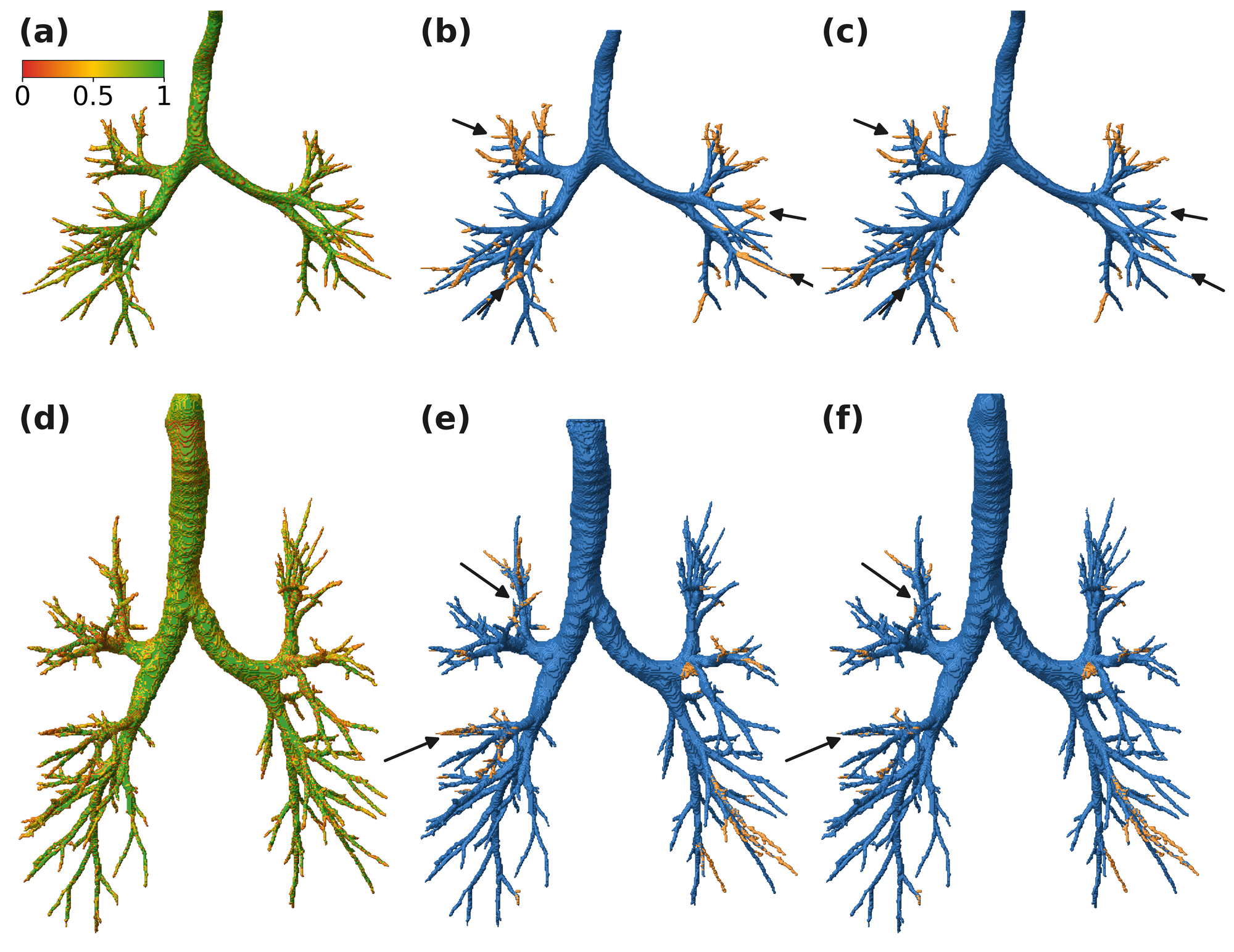}
\caption{A 1.7-year-old child (top) and an adult ATM test case (bottom). (a, d) \afuse{} segmentation $F$ colored by vote-pattern confidence (red low, green high). (b, e) \amr{} and (c, f) \afuse{} ($F$) in blue, with missed ground truth branches ($<80\%$ centerline coverage) in orange.  Arrows mark branches recovered by \afuse{}. Missed branches decrease from 109 to 66 of 292 in the child and from 50 to 36 of 293 in the adult; BND increases from 62.7\% to 77.4\% and 82.9\% to 87.7\%, respectively. This illustrates improved peripheral branch recovery in pediatric and adult CT.}
\label{fig:qualitative}
\end{figure}

\afuse{} has the highest BND at every adult generation, tying at 100\% in generations 1--3 (Table~\ref{tab:generation}). It reaches 90.95\% at generation 7 and 84.35\% at generation 10, versus 89.87\% and 82.42\% for \amr{}. Also, it has the highest clDice in generations 7--10. At generation 10, however, its DSC is slightly lower than that of \amr{}  (75.27\% vs.\ 75.74\%)  despite higher sensitivity (89.66\% vs.\ 88.66\%), reflecting additional  voxels outside the ground truth along thin branches. Thus, fusion primarily improves branch coverage rather than voxel overlap. Fig.~\ref{fig:qualitative} shows this for a child and an adult: \afuse{} recovers 43 and 14 of the branches missed by \amr{} alone.

\subsection{Supporting adult confidence analysis}
Vote-pattern confidence maps each model's prediction pattern to its voxel-level precision on the 30 validation volumes and is fixed before testing. On the 30 adult test volumes, it distinguishes false- from true-positive voxels in $F$ with a pooled AUROC of 0.891 (95\% CI 0.85--0.93, bootstrap over patients). The lowest-confidence 20\% of voxels contain 85\% of false positives (Fig.~\ref{fig:qualitative}a,~d).

\section{Discussion and Conclusion}
\afuse{} exploits complementary branches recovered by different models while avoiding disconnected false-positive components. Its anchor-based selection preserves connected branches from other models and avoids failures of largest-component post-processing. A second novelty is pediatric generalization without pediatric training: \afuse{} achieved the best performance across all six pediatric metrics and retained the airway in every examination, including cases where individual models failed. Limitations include the small single-center pediatric cohort, inability to remove false positives connected to the airway, dependence on \amr{} identifying the airway component and lack of independent validation of branch measurements. In summary, \afuse{} provides an anatomy-anchored strategy for robust adult and pediatric airway segmentation. To our knowledge, no available open-source method with pretrained weights is directly usable for pediatric CT, leaving an important gap for studies of airway development and pediatric airway disease. \afuse{} addresses this with a fully open pipeline including trained models, a web application and an ITK-SNAP 3D editing extension. Larger multicenter pediatric cohorts are needed for validation.

\section{Compliance with Ethical Standards}
Pediatric CT data were used under an IRB-approved protocol of Boston Children's Hospital with waiver of consent. Adult data are publicly released ATM challenge volumes, used under the challenge data usage agreement.

\section{Acknowledgments}
No funding was received. The authors declare no conflicts of interest. Claude and ChatGPT assisted with code and text editing. The authors verified all content. We thank Anish Misro, Alya Su Bolelli and Lara Su Bolelli for software testing.

\bibliographystyle{splncs04}
\bibliography{references}
\end{document}